\documentclass[a4paper]{PoS}

\usepackage{amsmath}
\usepackage{wasysym}
\usepackage{multirow}
\title{Hyperfine splitting in the bottomonium system 
on the lattice and in the continuum}

\ShortTitle{Hyperfine splitting in the bottomonium system 
on the lattice and in the continuum}

\author{\speaker{Nikolai Zerf}\\
                 University of Heidelberg - Institute for Theoretical Physics (ITP)\\
                 Philosophenweg 16, 69120 Heidelberg, Germany\\
          \thanks{This talk is based on Ref.~\cite{Baker:2015xma}.}\\
        E-mail:        \email{zerf@thphys.uni-heidelberg.de}}

\abstract{The latest experimental measurements of the hyperfine splitting
$E_{\rm{hfs}}=M_{\Upsilon(1 S)}-M_{\eta_b(1 S)}$
done by the Belle collaboration 
and the perturbative and lattice QCD predictions show a tension with the current extraction within lattice non-relativistic QCD.
In this talk we revise the analysis of radiative corrections to the bottomonium hyperfine splitting within lattice non-relativistic QCD at the next-to-leading order.
The result of the analysis is in good agreement with the perturbative and lattice QCD result as well as the direct measurement.
}

\FullConference{12th International Symposium on Radiative Corrections (Radcor 2015) and LoopFest XIV (Radiative Corrections for the LHC and Future Colliders)\\
 		15-19 June  2015\\
 		UCLA Department of Physics \& Astronomy Los Angeles, CA, USA}

\begin{document}

\section{Introduction}
The hyperfine splitting (HFS)
$E_{\rm{hfs}}=M_{\Upsilon(1 S)}-M_{\eta_b(1 S)}$
in the bottomonium system results within the Standard Model (SM) from a different spin configuration of the bottom quark and antiquark.
$M_{\Upsilon(1 S)}$ is a spin one (triplet) state where $M_{\eta_b(1 S)}$  is a spin zero (singlet) state.
A deviation of SM based predictions from the experimental results could be a hint of new physics and it is thus important to obtain an accurate SM based prediction.
Several distinct methods to predict $E_{\rm{hfs}}$ are being in use.
We shortly discuss a selection of them including the experimental measurements done so far in the respective overview section.
Then we revise the method employing lattice non-relativistic QCD (lNRQCD).
We present details required for the perturbative matching of QCD to lNRQCD at the one-loop level
and present our result.

\section{Experimental overview}
The $\Upsilon$ bound state in the spin one configuration was first observed in 1977 by the E288 collaboration in the di-muon channel as an excess over the Drell-Yan background in the proton beam collision with a fixed target.
The precise position of the $\Upsilon(1 S)$ resonance is nowadays measured with $e^+e^-$ colliders and the current PDG value is $\sqrt{s}=M(\Upsilon_{1S})=9460.30\pm0.26~\rm{MeV}$.

The $\Upsilon$ bound state in a spin zero configuration ($\eta_b$) was first observed by the Babar collaboration in 2008~\cite{Aubert:2008ba}.
Here $\eta_b(1 S)$ was produced in the radiative decay of $\Upsilon(3 S)$ plus an additional photon.
After the subtraction of the non-resonant background and the resonant background of the alternative decay into $\chi_b(2P)$ plus photon and the subtraction of the production of $\Upsilon(1 S)$ plus initial state radiation
the detected photon spectrum is peaked at the energy corresponding to a HFS of $E_{\rm{hfs}} = 71.4^{+2.3}_{-3.1}\pm2.7~\rm{MeV}\,$.
In 2009 Babar published another analysis using the radiative decay of $\Upsilon(2 S)$ getting a slightly lower value~\cite{Aubert:2009as}.
In 2010 the CLEO collaboration published an analysis based on the $\Upsilon(3 S)$ decays indeed finding an excess of events in the region where Babar found the $\eta_b$~\cite{Bonvicini:2009hs}.
However, the evidence of the excess is -- due to the small number of signal events -- less significant than the one from Babar.

In 2012 the Belle collaboration used a different production channel to determine the mass of $\eta_b$ with -- up to now -- highest precision~\cite{Mizuk:2012pb}.
Starting with a $\Upsilon(5 S)$ state decaying via an intermediate resonance into two pions and $h_b(1P)$ or $h_b(2P)$, 
which then decays further into $\eta_b(1S)$ plus photon, Belle finds consistent peaks in the $h_b(1P)$ and $h_b(2P)$ yield at an energy corresponding to a smaller HFS of $E_{\rm{hfs}} = 57.9 \pm 2.3 ~\rm{MeV}$.

The current world average by PDG~\cite{Agashe:2014kda} for the HFS uses was obtained by a weighted average of all four results leading a value of $E_{\rm{hfs}} = 62.3\pm 3.2~\rm{MeV}$.

\begin{figure}
\centering
\includegraphics[scale=0.5]{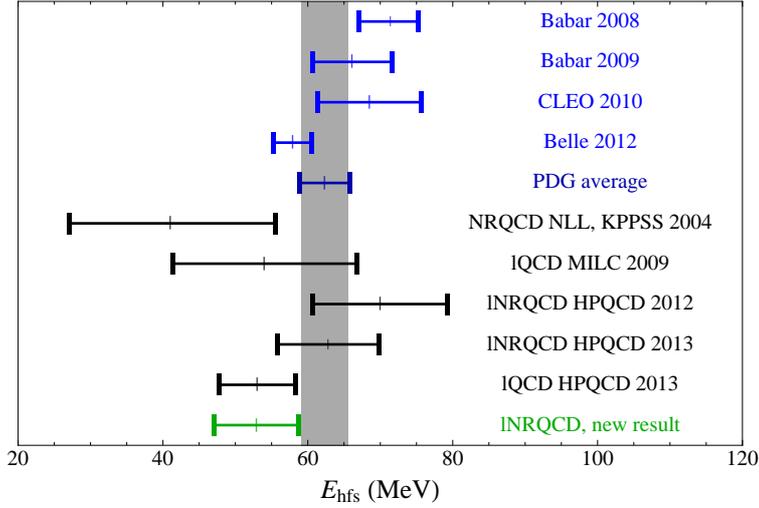}
\caption{Experimental and selected theory prediction values for the HFS.\label{FIG:HFSSummeryPlot} The gray band represents the world average determined by PDG.}
\end{figure}

All experimental values for the HFS are shown in Fig~\ref{FIG:HFSSummeryPlot}.

\section{Theory overview}
The following methods allow predictions for $E_{\rm{hfs}}$:
\begin{itemize}
\item{Phenomenologic potential models \cite{Ebert:2002pp}}
\item{Perturbative (potential) non-relativistic QCD (pNRQCD) calculation \cite{Kniehl:2003ap}}
\item{Lattice QCD simulations \cite{Burch:2009az,Davies:2013dem}}
\item{Lattice NRQCD simulations \cite{Dowdall:2011wh,Dowdall:2013jqa}}
\end{itemize}
The reference included in the list above lead to publications using the very method to obtain a value for $E_{\rm{hfs}}=M_{\Upsilon(1 S)}-M_{\eta_b(1 S)}$.

In the method of phenomenologic potentials one models a generic $q\overline{q}$-potential with free parameters in a first step.
Then the free parameters are fitted with the requirement that the model predictions are in agreement with experimental measurements for different mesonic states.
With the obtained fitted parameters the potentials are applied to the $b\overline{b}$ state with proper quantum numbers, 
where the energy shifts due to the HFS are evaluated within the framework of perturbation theory within quantum mechanics
involving wave-functions and Hamiltonians.
Although the obtained predictions can agree very well with the experimental measurements, 
the method itself is unable to judge if physics beyond the SM is hidden inside the experimental data,
because the fitted potentials may already contain such a contribution.
Moreover radiative corrections cannot be implemented in a systematic way following first principles.

In a perturbative pNRQCD calculation there is a systematic way to implemented radiative corrections within the tower of effective field theories QCD $\supset$ NRQCD $\supset$ pNRQCD.
Moreover, one can be sure that the prediction is purely based on QCD.
In Ref.~\cite{Kniehl:2003ap} $E_{\rm{hfs}}$ was calculated within the next-to-leading logarithmic approximation.
The uncertainty in the scale variations is still quite sizeable but the central value is lower than in other extraction method.
This can be due to missing non-perturbative contributions. However, there is no reliable estimate of the size of such contributions.

To cover all non-perturbative contributions one can rely on full lattice QCD simulations.
The difficulty here is that the bottom quark mass $m_b$ is much larger than $\Lambda_{QCD}$ and in order to resolve the UV dynamics of bottom quarks,
one is in principle forced to use a small lattice spacing $a\sim 1/m_b$.
At the same time one needs a large lattice to avoid cutting off non-perturbative IR dynamics $L \sim 1/\Lambda_{QCD}$.
Both requirements together lead to a computing intensive simulation setup which can be dealt with nowadays~\cite{Burch:2009az,Davies:2013dem}.  

To overcome the need to resolve the full dynamics up to scaled including $m_b$ one can just simulate the dynamics of NRQCD on the lattice.
Because NRQCD does not contain scales of the order of $m_b$ in dynamical fashion, one can chose the inverse lattice distance to be smaller than $m_b$.
In fact one requires
\begin{equation}
m_b > a^{-1} > m_b v\,.
\end{equation}
Here $v$ is the relative velocity of the bottom quark and antiquark,
and we require the inverse of $a$ to be larger than $m_b v$ because scales of this magnitude are still dynamical in NRQCD and thus need to be resolved in a lattice simulation.
The residual dependence of NRQCD on the hard modes $\sim m_b$ (which are not dynamical in NRQCD) is obtained via a matching calculation where one requires the agreement of QCD amplitudes with their NRQCD counterparts within the non-relativistic kinematics.
During this procedure free parameters in the NRQCD Lagrangian -- so called matching coefficients -- are determined.
Because QCD is perturbative at the scale $m_b$ one can obtain the matching coefficient within perturbation theory.

In the following we present the calculation of radiative corrections to the lNRQCD matching coefficient of the four fermion operator contribution to the HFS at the one loop order.

\section{Matching lNRQCD to QCD}
The terms responsible for generation of the HFS (at order order ${\cal O}(v^4)$) in the NRQCD
Lagrangian read  (see {\it e.g.}\cite{Pineda:1998kj,Pineda:1998kn})
\begin{equation}
{\cal L}_{\sigma}={c_F\over 2m_q}\psi^\dagger
{\bf B}{\bf \sigma}\psi +(\psi\to\chi_c)
+d_\sigma {C_F\alpha_s\over m_q^2}\psi^\dagger {\bf \sigma}\psi \chi_c^\dagger
{\bf \sigma}\chi_c,
\end{equation}
$\psi$ and $\chi_c$ are the two component Pauli-spinor fields of the heavy quark and anti-quark.
${\bf B}$ is the chromomagnetic field.
The relevant matching coefficients are the anomalous chromomagnetic coupling $c_F$ and the effective coupling constant $d_\sigma$ of the four fermion interaction.
$c_F$ can be determined non-perturbatively \cite{Meinel:2010pv,Dowdall:2011wh}
and is in good agreement with the perturbative one loop evaluation \cite{Hammant:2011bt}.
$d_\sigma$ was calculated in Ref. \cite{Hammant:2011bt}, 
however we do not agree with the obtained QCD amplitude.

In Fig we show the box diagrams required for the extraction of the matching coefficient $d_\sigma$.
\begin{figure}[t]
\centering
\begin{tabular}{ccc}
\includegraphics[scale=0.2,trim={0.0cm 5.0cm 0.0cm 5.0cm},clip]{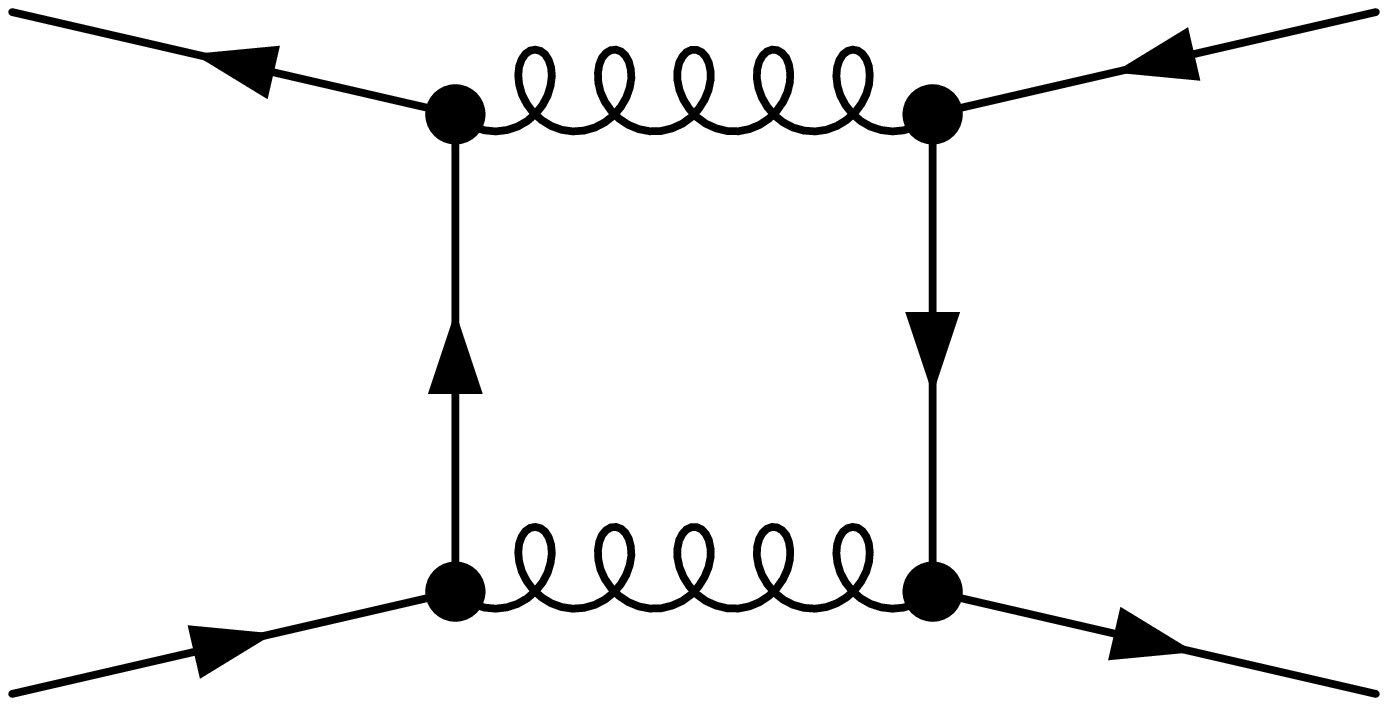}&
\includegraphics[scale=0.2,trim={0.0cm 5.0cm 0.0cm 5.0cm},clip]{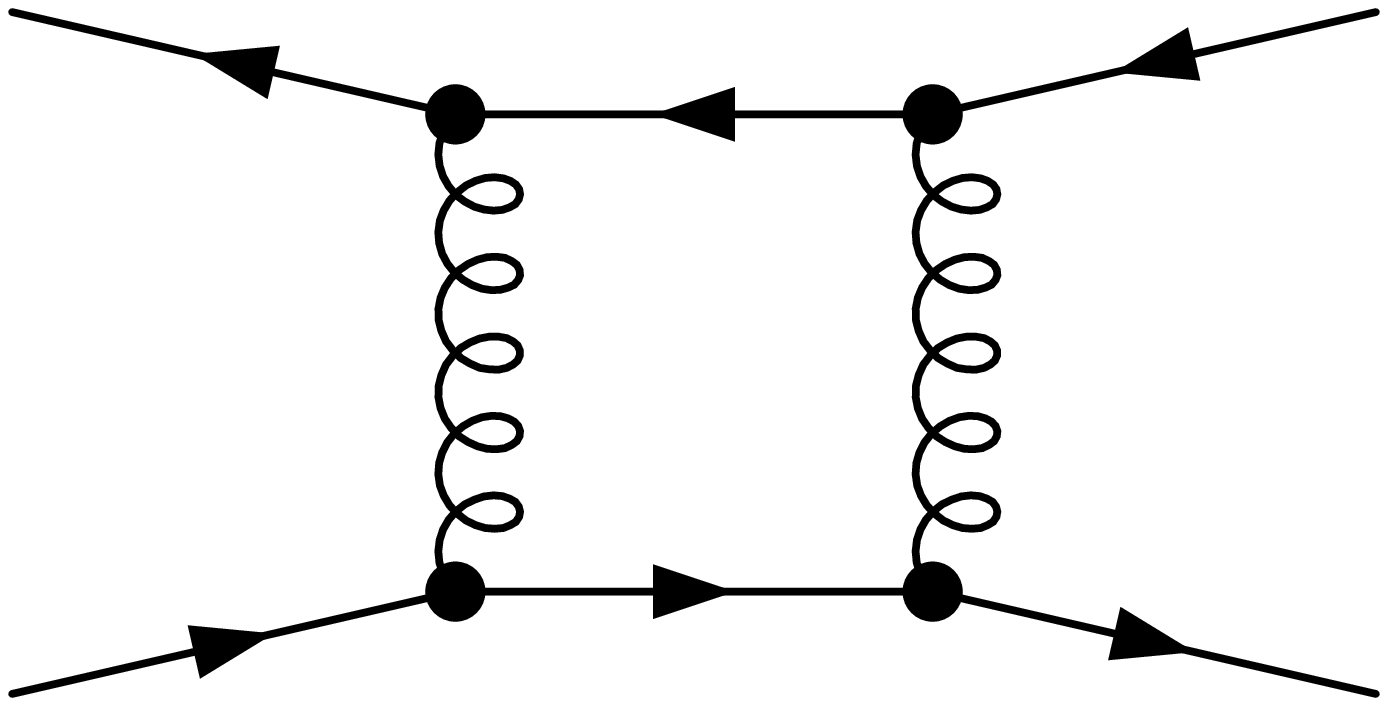}&
\includegraphics[scale=0.2,trim={0.0cm 5.0cm 0.0cm 5.0cm},clip]{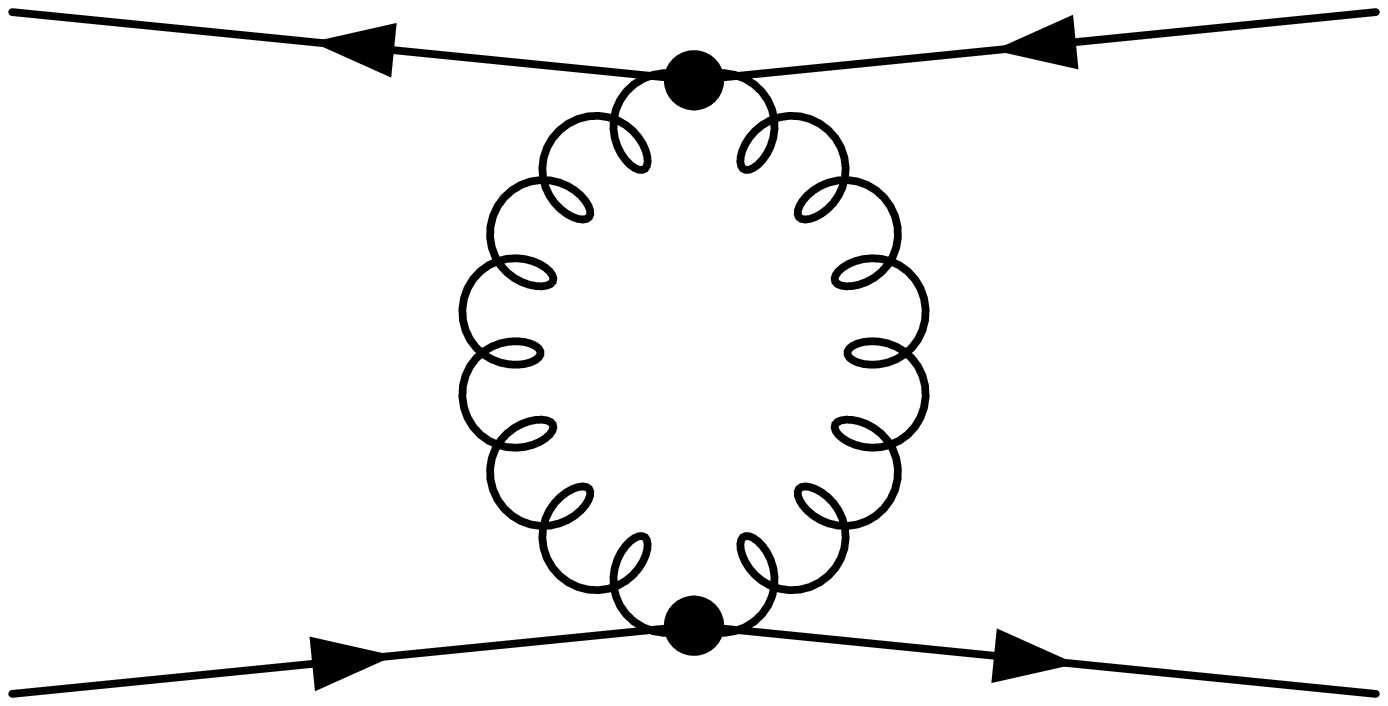}\\
(a)& (c) & (e)\\
\includegraphics[scale=0.2,trim={0.0cm 5.0cm 0.0cm 5.0cm},clip]{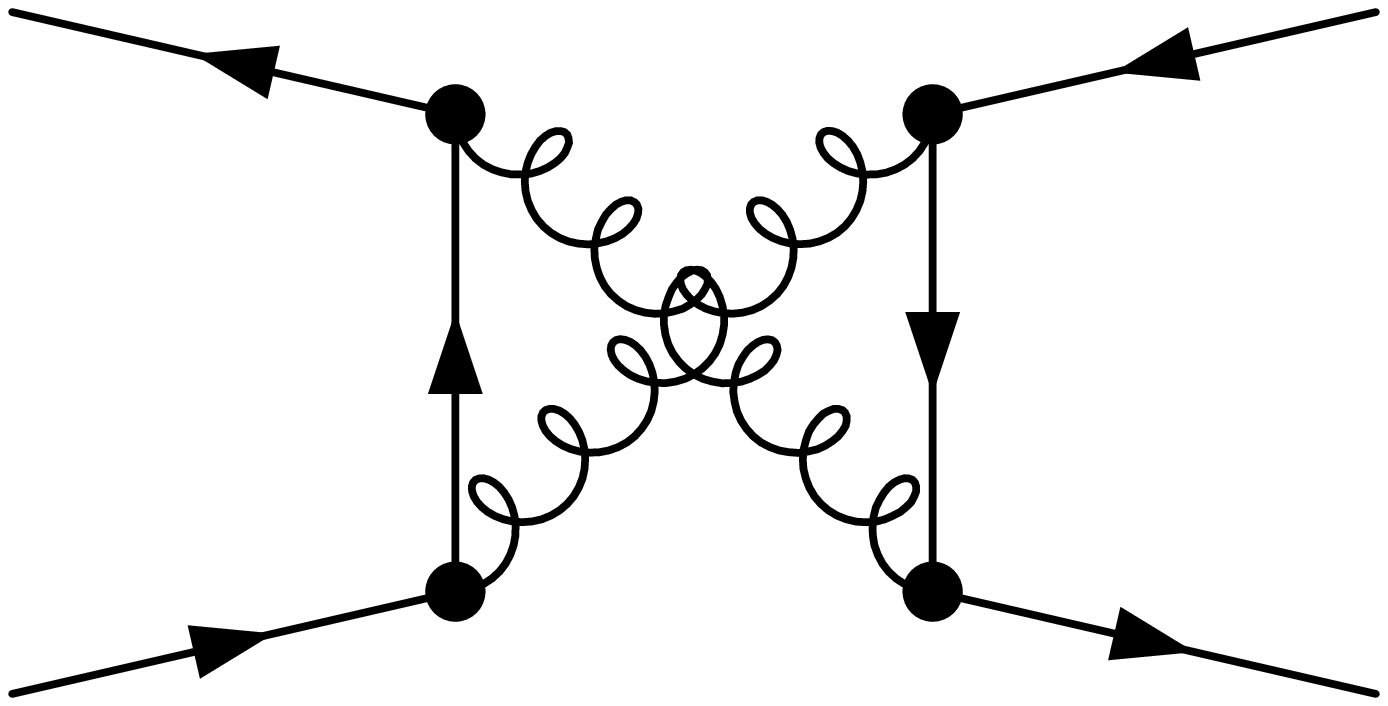}&
\includegraphics[scale=0.2,trim={0.0cm 5.0cm 0.0cm 5.0cm},clip]{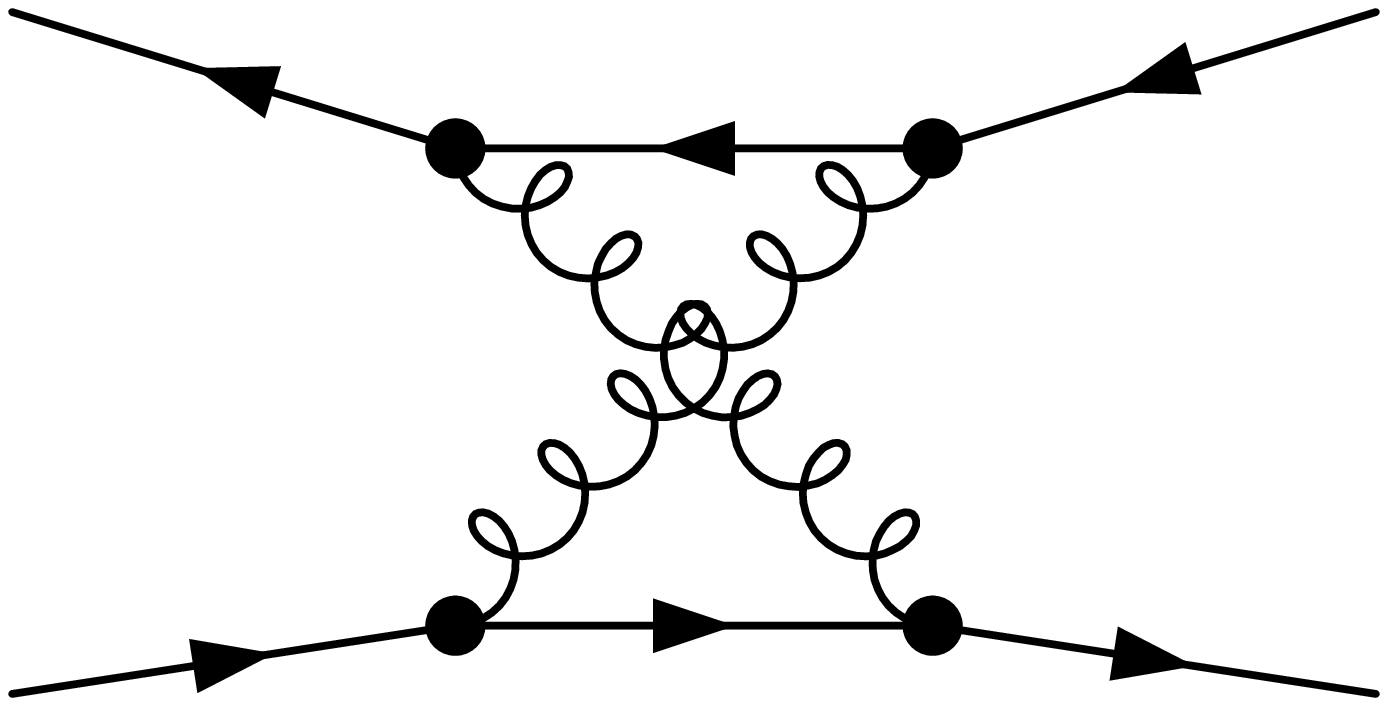}&
\includegraphics[scale=0.2,trim={0.0cm 5.0cm 0.0cm 5.0cm},clip]{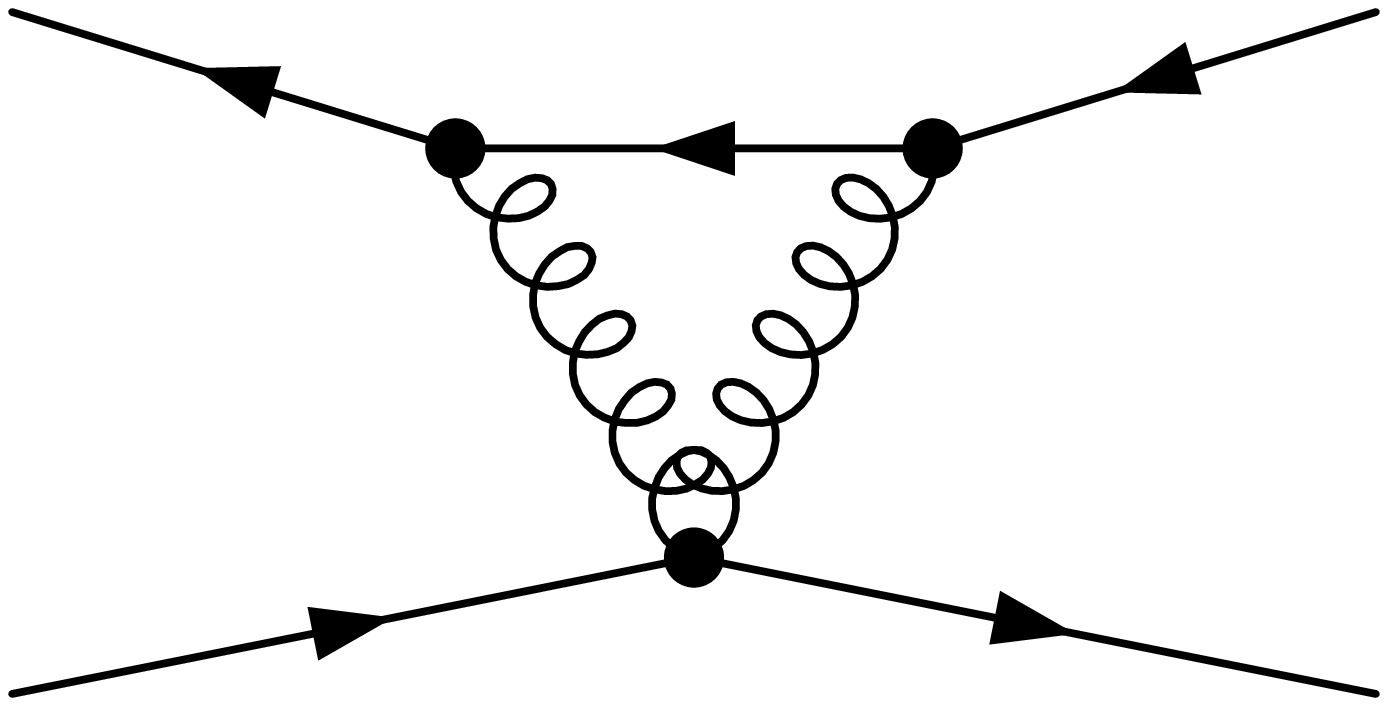}\\
(b)&  (d) & (f)
\end{tabular}
\caption{\label{fig::fig1}  One-loop box Feynman diagrams
relevant for the determination of $d_\sigma$.}
\end{figure}
In QCD diagrams (a-d) and in NRQCD diagrams (c-f) do contribute.
In order to regulate the corresponding amplitudes in the infrared we introduce a non-vanishing gluon mass $\lambda$.
Our QCD result reads:

\begin{eqnarray}
 M_{\rm 1PI}^{\rm QCD} &=&
\frac{C_F\alpha_s^2}{m_q^2}\left[{C_A\over 2}\log\left({m_q\over\lambda}\right)
+\left(\ln 2-1\right)T_F\right.
\nonumber\\&+&\left.\left(1-\frac{2\pi m_q}{3\lambda}\right)C_F
      \right]\psi^\dagger {\bf \sigma}\psi \chi_c^\dagger {\bf \sigma}\chi_c,
\nonumber \\
\label{eq::ampqcd}
\end{eqnarray}
Where we use $C_F=4/3, T_F=1/2$ and $C_A=3$.
Our lNRQCD amplitude reads:
\begin{eqnarray}
M_{\rm 1PI}^{\rm NRQCD} &=&
\frac{C_F\alpha_s^2}{m_q^2}\left[-\left(\delta
+{1\over 2}\ln\left(\lambda a\right)\right)C_A
\right.
\nonumber \\
&-&\left.\frac{2\pi m_q}{3\lambda}C_F\right]\psi^\dagger {\bf \sigma}\psi
\chi_c^\dagger {\bf \sigma}\chi_c
+{\cal O}\left(a^2\right),
\label{eq::ampnrqcd}
\end{eqnarray}
Besides a single negative power of $\lambda$ in the Abelian part ($\sim C_F$) corresponding to the Coulomb singularity 
there is only a logarithmic dependence on $\lambda$ up to higher order terms in $a\lambda$ which do vanish in continuum limit.
Because the infrared structure of lNRQCD and QCD do agree, all singularities appearing for $\lambda\rightarrow 0$ cancel in the difference of QCD and lNRQCD amplitude.
The non-singular Abelian part of the lNRQCD result vanishes in the limit $a\to 0$. Thus one only 
needs the lattice calculation of the non-Abelian term $\delta$. Note that beside the $a^2\lambda^2$ 
terms  the Abelian part of Eq.~(\ref{eq::ampnrqcd}) includes the linear lattice artifact $\sim a m_b$ associated with 
the lattice cutoff of the Coulomb singularity.

The difference of the QCD and lNRQCD amplitude has to be covered by an effective four-quark operator.
The coefficient in the Lagrangian is then given by
\begin{equation}
d_\sigma=\alpha_s\left[\left(\delta +{1\over 2}L\right)C_A
+\left(\ln 2-1\right)T_F+C_F\right],
\label{eq::ds}
\end{equation}
where  $L=\ln(m_q a)$.

In the case of a naively discretized lattice action an analytical calculation of the constant including the logarithm
using the expansion by region method introduce by Becher and Melnikov~\cite{Becher:2002if} was performed.
The analytic results reads: 
\begin{equation}
\delta^{\rm naive}=-{7\over 3}+28\pi^2 b_2-256\pi^2 b_3=0.288972\ldots,
\label{eq::delnaive}
\end{equation}
Here  the lattice tad-pole integrals are $b_2 = 0.02401318\ldots$, $b_3 = 0.00158857\ldots$.
In the calculation of the lNRQCD amplitude we systematically neglect higher-order 1/mb terms, 
in particular coming from the expansion of the non-relativistic quark propagator which we take in the static quark approximation.

We cross-checked our analytic result with an independent numerical calculation based on HiPPy/HPsrc package~\cite{Hart:2009nr}.
Here we use the program COLOR~\cite{vanRitbergen:1998pn} to reduce the color amplitudes analytically, leaving only the Lorentz-amplitudes for the integration with the CUBA library~\cite{Hahn:2004fe}.

For the improved lattice action  -- used by the HPQCD collaboration~\cite{Dowdall:2011wh}  in lattice simulation -- we use the same numerical setup, 
except we provide HPsrc with the electronic Feynman-rules following from the improved action.
Our numerical result for the improved action reads then:
\begin{equation}
\delta=0.1446(28)\,.
\label{eq::delHPQCD}
\end{equation}

\section{Determination of the HFS}
The HFS was calculated within a lattice simulation for three different lattice spacings.
First taking into account terms up to ${\cal O}(v^4)$ in the action action~\cite{Dowdall:2011wh} and later including ${\cal O}(v^6)$  terms~\cite{Dowdall:2013jqa}.
The lattice result without the four-quark operator contribution is available inside the stated references.
In order to take the four quark operator contributing to the HFS into account we add its effect
\begin{equation}
\Delta E_{\rm hfs}=-d_\sigma{4 C_F\alpha_s\over m_q^2}|\psi(0)|^2,
\label{eq::DelE}
\end{equation}
on top of the lattice data for the HFS.
For the ${\cal O}(v^4)$ and ${\cal O}(v^6)$ action we obtain from a constrained fit:
\begin{align}
 E_{\rm hfs}^{{\cal O}(v^4)}=57.5\pm_{\text{dis}}2.6\pm_{\text{rel}}6.0\pm_{\text{rad}}4.8~\text{MeV}\,,\nonumber\\
 E_{\rm hfs}^{{\cal O}(v^6)}=51.5\pm_{\text{dis}}3.1\pm_{\text{rel}}1.8\pm_{\text{rad}}4.3~\text{MeV}\,.
\end{align}
Here the first error indicates the discretization uncertainty.
The second error is due to neglected higher order relativistic corrections 
and the last error reflects the uncertainty of higher order radiative correction. 
The latter is estimated via the size of the known double logarithmic two loop terms of $c_F$ and $d_\sigma$ as well as the stated numerical uncertainty in the one loop coefficient $c_F$.
We combine the two results where we only treat the error due to higher order radiative corrections correlated.
The final result reads:
\begin{align}
 E_{\rm hfs}=52.9\pm5.5~\text{MeV}\,.
\end{align}

\section{Discussion}
In Fig.~\ref{FIG:HFSSummeryPlot} we compare our result with all experimental determinations and a selection of previous theory determinations.
Compared to the most recent lNRQCD analysis value our result is about $10~\text{MeV}$ smaller.
Half of the difference is due to the difference in the one-loop QCD amplitude 
(the QCD result  of Ref.~\cite{Hammant:2011bt} has been corrected shortly after this talk, see the erratum, and now agrees with eq. (\ref{eq::ampqcd})).
The remaining difference is mainly a consequence of the different treatment of lattice artifacts and $1/(am_b)$ suppressed contributions in $d_{\sigma}$.
We do not include any lattice artifacts proportional to $a m_b$ and/or $1/(am_b)$ suppressed contributions in the matching coefficient.
The $1/(am_b)$ contributions are small for lattice spacings used in the actual simulations.
At the same time the one-loop contribution does not account for the linear lattice artifact in 
the given order. Due to the non-perturbative character of the Coulombic interaction 
the contribution of the multiple Coulomb gluon exchanges to the coefficient of  $\sim a m_b$ 
term is not suppressed. The latter should be determined  from the non-perturbative lattice data 
and the one-loop result can only be used for a rough estimate. We have found that  the  effect of 
the linear artifact on the continuum extrapolation is within the error bars given above.
For the ${\cal O}(v^4)$ action the continuum limit works extremely well.
We expect that for the ${\cal O}(v^6)$ action the continuum limit improves after taking into account the one loop matching for all relevant higher order operators up to order ${\cal O}(v^6)$.

\acknowledgments
The author would like to thank the organizers for the nice conference at UCLA.
Further thanks go to Alexander Penin for many useful discussions and reading the manuscript.


\begin{thebibliography}{99}

\bibitem{Baker:2015xma}
  M.~Baker, A.~A.~Penin, D.~Seidel and N.~Zerf,
  Phys.\ Rev.\ D {\bf 92} (2015) 054502
  [arXiv:1504.05979 [hep-ph]].

\bibitem{Aubert:2008ba}
  B.~Aubert {\it et al.}  [BaBar Collaboration],
  Phys.\ Rev.\ Lett.\  {\bf 101}, 071801 (2008)
  [Erratum-ibid.\  {\bf 102}, 029901 (2009)].

\bibitem{Aubert:2009as}
  B.~Aubert {\it et al.}  [BaBar Collaboration],
  Phys.\ Rev.\ Lett.\  {\bf 103}, 161801 (2009).

\bibitem{Bonvicini:2009hs}
  G.~Bonvicini {\it et al.}  [CLEO Collaboration],
  Phys.\ Rev.\ D {\bf 81}, 031104 (2010).

\bibitem{Mizuk:2012pb}
  R.~Mizuk {\it et al.}  [Belle Collaboration],
  Phys.\ Rev.\ Lett.\  {\bf 109}, 232002 (2012).

\bibitem{Agashe:2014kda}
  K.~A.~Olive {\it et al.} [Particle Data Group Collaboration],
  Chin.\ Phys.\ C {\bf 38} (2014) 090001.

\bibitem{Ebert:2002pp}
  D.~Ebert, R.~N.~Faustov and V.~O.~Galkin,
  Phys.\ Rev.\ D {\bf 67} (2003) 014027
  [hep-ph/0210381].

\bibitem{Kniehl:2003ap}
  B.~A.~Kniehl, A.~A.~Penin, A.~Pineda, V.~A.~Smirnov and M.~Steinhauser,
  Phys.\ Rev.\ Lett.\  {\bf 92}, 242001 (2004).

\bibitem{Burch:2009az}
  T.~Burch, C.~DeTar, M.~Di Pierro, A.~X.~El-Khadra, E.~D.~Freeland, S.~Gottlieb, A.~S.~Kronfeld and L.~Levkova {\it et al.},
  Phys.\ Rev.\ D {\bf 81}, 034508 (2010).

\bibitem{Davies:2013dem}
  C.~T.~H.~Davies {\it et al.} [HPQCD Collaboration],
  PoS LATTICE {\bf 2013} (2014) 438
  [arXiv:1312.5874 [hep-lat]].

\bibitem{Dowdall:2011wh}
  R.~J.~Dowdall {\it et al.}  [HPQCD Collaboration],
  Phys.\ Rev.\ D {\bf 85}, 054509 (2012).

\bibitem{Dowdall:2013jqa}
  R.~J.~Dowdall {\it et al.}  [HPQCD Collaboration],
  Phys.\ Rev.\ D {\bf 89}, no. 3, 031502 (2014).

\bibitem{Pineda:1998kj}
  A.~Pineda and J.~Soto,
  Phys.\ Rev.\ D {\bf 58}, 114011 (1998).

\bibitem{Pineda:1998kn}
  A.~Pineda and J.~Soto,
  Phys.\ Rev.\ D {\bf 59}, 016005 (1999).

\bibitem{Meinel:2010pv}
  S.~Meinel,
  Phys.\ Rev.\ D {\bf 82}, 114502 (2010).

\bibitem{Hammant:2011bt}
  T.~C.~Hammant, A.~G.~Hart, G.~M.~von Hippel, R.~R.~Horgan and C.~J.~Monahan,
  Phys.\ Rev.\ Lett.\  {\bf 107} (2011) 112002
   [Phys.\ Rev.\ Lett.\  {\bf 115} (2015) 039901]
  [arXiv:1105.5309 [hep-lat]].

\bibitem{Becher:2002if}
  T.~Becher and K.~Melnikov,
  Phys.\ Rev.\ D {\bf 66}, 074508 (2002).

\bibitem{Hart:2009nr}
  A.~Hart, G.~M.~von Hippel, R.~R.~Horgan and E.~H.~Muller,
  Comput.\ Phys.\ Commun.\  {\bf 180}, 2698 (2009).

\bibitem{vanRitbergen:1998pn}
  T.~van Ritbergen, A.~N.~Schellekens and J.~A.~M.~Vermaseren,
  Int.\ J.\ Mod.\ Phys.\ A {\bf 14} (1999) 41.

\bibitem{Hahn:2004fe}
  T.~Hahn,
  Comput.\ Phys.\ Commun.\  {\bf 168} (2005) 78.

  
\end{thebibliography}
\end{document}